\documentclass[12pt]{article}  

\usepackage{latexsym}
\usepackage{graphicx}
\usepackage{amsfonts}
\usepackage{amssymb}
\usepackage{amsmath}
\usepackage{hyperref}

\usepackage{multirow}

\def\be{\begin{equation}}
\def\ee{\end{equation}}
\def\ba{\begin{eqnarray}}
\def\ea{\end{eqnarray}}
\def\la{\label}

\newcommand{\beq}{\begin{equation}}
\newcommand{\eeq}[1]{\label{#1}\end{equation}}
\newcommand{\bea}{\begin{eqnarray}}
\newcommand{\eea}[1]{\label{#1}\end{eqnarray}}

\newcommand{\sst}[1]{{\scriptscriptstyle #1}}

\def\0{{\sst{(0)}}}
\def\1{{\sst{(1)}}}
\def\2{{\sst{(2)}}}
\def\3{{\sst{(3)}}}
\def\4{{\sst{(4)}}}
\def\5{{\sst{(5)}}}
\def\6{{\sst{(6)}}}
\def\7{{\sst{(7)}}}

\begin{document}
\begin{flushright}
\end{flushright}

\vspace{20mm}

\begin{center}

{\bf Montonen-Olive duality of gauged supergravity?}
\footnote{To be submitted for publication in {\it Fields, Gravity, Strings and Beyond:}\\In Memory of Stanley Deser, J.Phys.A . Eds. Henneaux, Nepomechie and Seminara} 
\vspace*{0.37truein}
\vspace*{0.15truein}

M.~J.~Duff
\footnote{m.duff@imperial.ac.uk} \\
\vspace*{0.15truein}
{\it Blackett Laboratory, Imperial College London} \\
{\it Prince Consort Road, London SW7 2AZ} \\


\baselineskip=10pt
\bigskip

\vspace{20pt}

\abstract{

\noindent The vanishing of the $SO(N)$ $\beta$-function of gauged $N>4$ supergravity has long seemed like  an answer looking for a question. Here we conjecture that it is a consequence of a Montonon-Olive style electric/magnetic duality.}

\vspace{20pt}

~~~~~~~~~~~~~~~~~~~~Dedicated to the memory of Stanley Deser
\end{center}
\newpage

\tableofcontents

\newpage


\section{Stanley Deser}

Fleeing 1930s Poland first for Palestine and then for France; escaping Paris in 1940 hours before the Nazi tanks came rolling in; catching the last safe train out of France with life-saving transit visas issued by an heroic Portuguese consul; securing passage on one of the last boats out of Lisbon to New York.  Anyone with a childhood as tumultuous as this might reasonably expect adulthood to be something of an anticlimax, but for Stanley Deser another adventure was only just beginning, albeit an intellectual one in theoretical physics. 
Readers of his autobiography {\it Forks In The Road} \cite{Deser:2023} can relish the anecdotes, told with wit and literary aplomb, involving the greatest minds of 20th century physics. 
They will come away with a greater appreciation of a golden era for quantum field theory and general relativity to which Deser was himself a major contributor. One such story recalls
 former hero of the Manhattan Project J Robert Oppenheimer, betrayed and humiliated by H-bomb advocate Edward Teller, phoning Teller to politely request he approve the young Deser's job application (which he did).

 Quantum Gravity is the would-be unified theory that eluded Einstein, mystified the greatest minds of twentieth century physics and continues to baffle those of the twenty first. Stanley Deser, who was Ancell Professor Emeritus at Brandeis University and Visiting Professor at Caltech, can justifiably claim not only to be one of the world's leading experts on this conundrum but through his work on the canonical formalism, supergravity and string theory perhaps also one of the providers of its solution. Building on earlier work of Dirac, the Arnowitt-Deser-Misner (ADM) formulation \cite{Arnowitt:1959ah} did for Einstein's general relativity what Hamilton did for classical mechanics. As such it also formed the basis of early attempts to quantize gravity and continues to be used in modern methods based on supergravity, extra dimensions, string and M-theory.  The discovery of Supergravity, a supersymmetric version of general relativity, by Freedman, Ferrara and Van Nieuwenhuizen  \cite{Freedman:1976xh} and by Deser and Zumino \cite{Deser:1976eh} represents an important milestone in theoretical physics and is probably Stanley's most well known paper. If Einstein had not already discovered general relativity, supersymmetry would have forced us to invent it:  a local supersymmetry transformation is the square root of a general co-ordinate transformation, the basis of relativity.
A major problem that plagues attempts to treat Einstein's theory as a quantum field theory is that of non-renormalizability and his work with Kay and Stelle \cite{Deser:1977yyz} showed how supersymmetry ameliorates the infinities he found in the non-supersymmetric case with Van Nieuwenhuizen \cite{Deser:1974cz} . It is still an open question whether the maximally supersymmetric supergravity is or is not finite.  For this and other reasons, theorists' attention turned to string theory to which Deser has also made seminal contributions  \cite{Deser:1976rb,Boulware:1985wk}. Supergravity continues to be an active area of research both as the low energy limit of string theory and in its own right.  Further work with Isham and me on quantum gravity  \cite{Deser:1976yx} revealed the existence of Weyl anomalies: the conformal invariance displayed  by classical massless field systems in interaction with gravity no longer survives in the quantum theory. These have found a variety of applications in unified theories of the elementary particles, in condensed matter physics, in the AdS/CFT correspondence and in cosmology. The classification of these anomalies in arbitrary dimensions may be found in his paper with Schwimmer  \cite{Deser:1993yx}. Other contributions include gravity in three dimensions \cite{Deser:1983tn}, broken supersymmetry in supergravity \cite{Deser:1977uq} and a proof of its positive energy \cite{Deser:1977hu}.  In the 1980s he was instrumental in securing NSF approval of the LIGO gravitational wave experiment against much opposition.  

He was awarded the Einstein Medal in 2015 for important contributions to general relativity, in particular the development of the ADM formalism. Other awards include: The Dannie Heineman Prize 1994; Fellow of the National Academy of Sciences 1994; Honorary Member of the Torino Academy of Sciences 2001; Member of the American Academy of Arts and Sciences 1979, Foreign Member of the Royal Society 2021.

 \begin{figure}\label{fig:p4}  
\begin{center}
\includegraphics[scale=0.7]{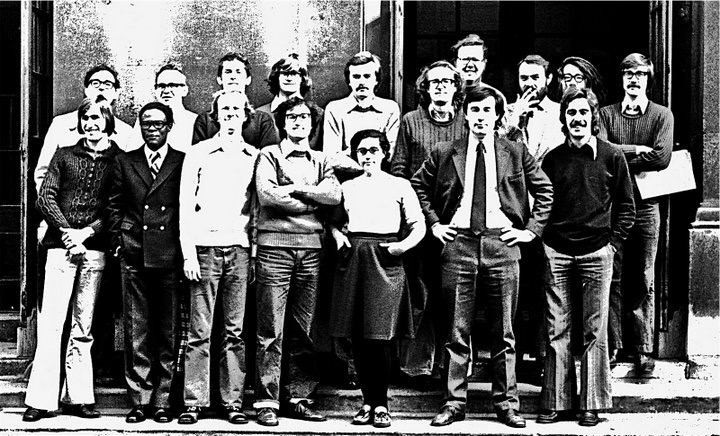}
\caption{\footnotesize{King's College London 1976. FRONT ROW: Dave Storey, Unknown, Steve Christensen, Bernard Kay, Joyce El Eini, Chris Isham, Mike Duff. BACK ROW: John Taylor, Steve Fulling, Stephen Downs-Martin, Tim Bunch, Chris Mannion, David Robinson, Bob Davis, Stanley Deser, Tezt Yoshimura, Paul Davies. }}
\end{center}
\end{figure}

I first came into contact with Stanley as a student at the last (1970) Brandeis Summer School, of which he was a co-organizer. He was easy to spot because there was always a cloud of pipe-smoke following him around, but I did not get to know him personalIy until he visited King's College London where he was a guest of Chris Isham and I was a postdoc. In fact, he had a long association with the UK, dating back to collaborations with Hermann Bondi and Felix Pirani at King's, with Dennis Sciama at Oxford where he was visiting Fellow of All Souls and Gary Gibbons at Cambridge. The combination of permanent staff, postdocs and visitors made King's a centre of activity for classical gravity (Felix Pirani and David Robinson), quantum field theory in curved spacetime (Tim Bunch, Steve Christensen, Paul  Davies, Larry Ford, Steve Fulling and Bill Unruh) and quantum gravity (Stanley Deser, Chris Isham, John G. Taylor and myself).  Many appear in the photograph Fig. \ref{fig:p4}, courtesy of Steve Christensen.  See \cite{Robinson:2018oax} for a history.  But the visitor who had the greatest impact on Chris and me was Stanley (the one smoking the cigar): non-local conformal anomalies \cite{Deser:1976yx}, solitons of chiral theories in three space dimensions \cite{Deser:1976qe}, sine-Gordon/Thirring equivalence in curved spacetime.  I have related the Weyl anomaly saga elsewhere \cite{Duff:1993wm,Duff:2020dqb}. The chiral solitons paper introduced fourth order terms in the lagrangian necessary to circumvent the Derrick no-go theorem in three dimensions\footnote{An alternative circumvention, teetering on the rim of the dustbin, was to raise the sigma-model lagrangian to the power 3/2.}.  We obtained what we considered to be nice results only to discover that some had been anticipated by Skyrme in the early 60s, although this is overlooked in the received history of  the Skrymion revival \cite{Aitchison:2019hzj}. As for the well-known equivalence of the sine-Gordon and Thirring models, we originally claimed that it failed in curved spacetime and submitted a paper for publication only to realize they were in fact equivalent. (If my memory serves me correctly, this involved some subtleties with the beta functions and the two-dimensional Weyl anomaly).  We issued a hasty retraction but, for reasons I cannot recall, nothing in the end got published. 

My interactions with Stanley were set to increase as I accepted his offer of a two-year NSF postdoc at Brandeis, where there was a very active  group including Marc Grisaru, Howie Schnitzer and Burt Ovrut. I also  enjoyed the warm hospitality extended by Stanley and his wife Elsbeth at their home in Newton. Elsbeth was an accomplished artist and daughter of the Swedish physicist Oskar Klein, whose work on extra dimensions was to have a profound effect on my career.

 Boston was, and still is, the place to be for theoretical physics. Moreover, by masquerading as Martin Rocek, who lent me his
Harvard ID card, I was able to get an apartment in Cambridge more easily and was thus able to interact with the people at Harvard (where I had desk space) such as (importantly for the present paper) my old friend from King's Steve Christensen. Looking back,  I think how lucky I was to be in Boston at that time both scientifically and socially. There was a group of us who went to the Chinese restaurant after the joint Harvard/MIT seminars: Stanley Deser, Roman Jackiw, Sydney Coleman and occasionally Geoffrey Goldstone. Lunch with Steve Weinberg at the Harvard Faculty Club was another privilege.  

See \cite{Christensen:2022} for Steve Christensen's nostalgic recollections of our Brandeis/Harvard collaboration which resulted in some of my most cited papers. Our 1978 paper on index theorems was the first to calculate spin 3/2 anomalies (and one of the first to use a word processor!) \cite{Christensen:1978gi,Christensen:1978md}.  We followed this with a paper on quantizing gravity with a cosmological constant \cite{Christensen:1979iy} calculating in particular how the cosmological constant, $\Lambda$, is renormalized by closed loops of spins 0, 1/2, 1 and 2.  To do the spin 3/2 case, which essentially means gauged  $ N \leq 8$ supergravity, we teamed up with Martin Rocek and Gary Gibbons. This  case is particularly interesting because the product of the cosmological constant and Newton's constant, G,  given by the $SO(N)$ coupling constant, e,
\[e^2~\sim -G\Lambda\]
and hence its renormalization is given by the $SO(N)$ beta function. We found that the beta function vanishes for $N>4$ \cite{Christensen:1980ee}. 
Given the importance of supergravity in Stanley's life and mine, this is the topic I have chosen to write about in this
dedication. Our conjecture will be that just as $N>2$ Yang Miils admits an exact electric/magnetic duality which demands  a vanishing beta, so does gauged $N>4$ supergravity.

Other memories  of Stanley include our stay in the Academy of Sciences Hotel for the Moscow Quantum Gravity conference circa late 1970s. Stanley and I set out one night for dinner at a nearby restaurant. In those bleak years before Glasnost it was not uncommon to be turned away from half-empty restaurants on the grounds that they were ``full''.  This is as exactly what happened, but Stanley Deser would not take no for an answer and stuck his foot in the door as the proprietor
tried to close it in our faces. Fisticuffs broke out until I managed to pull Stanley away arguing that the meal was probably not worth getting beaten up for.

Stanley passed away in 2023 but in the later stages of his illness we kept up a vigorous exchange of e-mails.  Although those who have corresponded with Stanley will recall that his messages, often crammed into the subject line,  sometimes needed an Enigma Machine to decipher his cryptic comments.  A formidable intellect, a cultured polymath, linguist, connoisseur of fine food and wine, {\it frivolous} is the last adjective that springs to mind when recalling Stanley
 Deser.  All the more enjoyable therefore was the game we played: {\it Physics-based Horror Movies.} This was prompted by an invitation I received from Peter Freund  to deliver the 9th annual Schrodinger lecture in Transylvania. We came up with some obvious ones to start with: {\it Frank Einstein}, {\it Mary Shelley Glashow}, {\it Fadeev-Popov Ghosts}, {\it Maxwell's Demon},  {\it Phantom Black Holes}, {\it The Bourne Approximation}\footnote{OK, not strictly horror.}; then some more mathematical ones: {\it Bram Stoker's Theorem}, {\it The Killing Vector}, {\it The Texas Chain Rule Massacre}, {\it Paranormal Coordinates}, but Stanley trumped them all with ...{\it Diracula}! 

\section{Vanishing of the $SO(N)$ $\beta$-function of gauged $N>4$ supergravity}

\begin{table}[h!]
$\begin{array}{llrrrrrrrrrrrrr}
s&&-60B(s)&&960\pi^2\beta(s)/e^3&\\
&&&&&&&\\\\
\bigskip
2&&522&&0&~&&\\
3/2&&-137&&26C(3/2)&\\\\
1&&12&&-11C(1)&\\\\
1/2&&3&&2C(1/2)&\\\\
0&&2&&C(0)&\\\
\end{array}$
\label{BandB}
\caption{The coefficents $B$ and $\beta $ for fields of spin $s$(counting the scalars as complex)}
\end{table}

\begin{table}[h!]
$\begin{array}{llrrrrrrrrrrrrr}
N&&B(total)&&-16\pi^2\beta(total)/e^3&\\
&&&&&&&\\
\bigskip
1&&-77/12&&-&~&&\\
2&&-13/3&&-13/3&&& \\\\
3&&-5/2&&-5/2&\\\\
4&&-1&&-1&&\\\\
5&&0&&0&&& \\\\
6&&0&&0&\\\\
7&&0&&0&&\\\\
8&&0&&0\\\\
\end{array}$
\label{matching}
\caption{The values of $B$ and $\beta$ in gauged $N$-extended supergravity, demonstrating their equivalence}
\end{table}

In 1980, Steve Christensen and I calculated the one-loop counterterms for pure gravity with a cosmological constant \cite{Christensen:1980ee}.
\be
S=-\frac{1}{\kappa^2}\int d^4x(R-2\Lambda)
\label{einstein1}
\ee
The trick was to work in coordinate space and to expand about a background satisfying the field equations\footnote{Even today I see papers on the cosmological constant getting themselves into trouble by  expanding around a flat background in order to work in momentum space}
 \be
R_{\mu\nu}=\Lambda g_{\mu\nu}
\ee
\la{einstein2}
We adopted dimensional regularisation with spacetime dimension $4+\epsilon$ .The one-loop counterterms will be of the form $R_{\mu\nu\rho\sigma}R^{\mu\nu\rho\sigma}$,  $R_{\mu\nu}R^{\mu\nu}$, $R^2$, $\Lambda R$, $\Lambda^2$ with gauge-dependent coefficients.  Gauge invariance is achieved by going ``on-shell'' by use of the field equations  (\ref{einstein2}). The result is
\be
\Delta S=\frac{1}{\epsilon}(A\chi-B\frac{\kappa^2 \Lambda}{12 \pi^2} S)
\label{counter1}
\ee
where  $A$ and $B$ are numerical coefficients, $\chi$ is the Euler number of spacetime
\be
\chi=\frac{1}{32\pi^2}\int d^4x \sqrt{-g}(R_{\mu\nu\rho\sigma}R^{\mu\nu\rho\sigma}-4R_{\mu\nu}R^{\mu\nu}+R^2)
\ee
and S is the classical action on shell. Explicit calculation yields
\be
A=\frac{106}{45},  ~~~~~~~~B=-\frac{87}{10}
\ee
Thus, pure gravity with a cosmological constant  is no longer one-loop finite (in the non-topological sense) because $B\neq 0$.  Next we included closed loop of matter 
particles with spins 0, 1/2, and 1 \cite{Christensen:1980ee} with the results shown in Table \ref{BandB}. 
The only consistent way to include spin 3/2 is to go to the gauged
supergravity  \cite{Freedman:1976aw}. Here the global $SO(N)$ of $N$-extended supergravity is gauged with gauge coupling $e$ given by
\be
\kappa^2\Lambda=-6e^2
\ee
    so the counterterm action (\ref{counter1}) becomes
\be
\Delta S=\frac{1}{\epsilon}(A\chi+B\frac{e^2 }{2 \pi^2} S)
\label{counter2}
\ee
where S now includes the Yang-Mills kinetic term. The beta function  
$\beta$ is given by
\be 
B=-\frac{16\pi^2\beta}{e^3} 
\label{equality}
\ee
and the signal for asymptotic freedom is $B>0$. A strong consistency check is now to calculate $\beta$ from the charge renormalization of an $SO(N) $ Yang-Mills gauge theory (something with which we were familiar) and show that it gives the same answer as cosmological constant renormalization for $N>1$ (something with which we were less familiar). See Table \ref{BandB} where C(s) are the quadratic Casimirs. Details may be found in \cite{Duff:1982yw}. Note that the equality (\ref{equality}) holds only for the total B and $\beta$ not spin by spin. For example the graviton, being electrically  neutral, does not contribute at all to $\beta$.  As shown in Table \ref{matching}, we did indeed find agreement between the total B and $\beta$ in N-extended supergravity. The breakdown to $N= 1$ through $N=8$ multiplets may be found in the Appendix.  For $N\leq4$, $\beta$ is positive and so gravitinos are not confined. For $N>4$, 
$\beta$ vanishes \footnote{When these calculations were carried out, gauged supergravity for $N>4$ had not yet been constructed. This required some educated guess-work on our part. I told Bruno Zumino our result, hoping to impress him but received the withering response ``Congratulations, you have shown beta is zero for a theory that does not exist''. Fortunately,  we were vindicated by de Wit and Nicolai \cite{deWit:1982bul} when they constructed the $N=5,6,8$ cases.}. Moreover, Stelle and Townsend claim that this is true not just to one loop but to all orders \cite{Stelle:1982ug}.

Another way to understand this is to note that in the arbitrary spin approach \cite{Christensen:1980ee}, $B$ is given by a fourth order polynomial 
\be
B(total)=\Sigma_s d(s)B(s)
\ee
\be
60B(s)=(-1)^{2s}(-2+30s^2-40s^4)
\ee
where $d(s)$ is the number of fields of spins $s$ in the supermultiplet (again counting complex scalars).
One may now invoke the supersymmetric spin sum rule \cite{Curtright:1981wv}
\be
\Sigma_s(-1)^{2s}d(s) s^k=0 ~~~~~~~N>k
\label{sum}
\ee
                In fact it was further conjectured in \cite{Duff:1982ev} and proved in  \cite{Gibbons:1984dg,Inami:1984vp,Duff:2002sm}, using zeta function methods, that the vanishing of $\beta(e)$ continues to hold  for the massive Kaluza-Klein tower arising from the round seven-sphere compactification of eleven dimensional supergravity \cite{Duff:1986hr}. Of course in Kaluza-Klein when an odd dimensional theory, which is one-loop finite, compactifies to an even dimensional theory whose zero-modes yield infinities, it must be that these infinities are cancelled by equal and opposite contributions from the massive modes \cite{Duff:1982gj}. This yields vanishing A in (\ref{counter2}), but interestingly B vanishes level by level in the KK tower \cite{Gibbons:1984dg,Inami:1984vp,Duff:2002sm}. 

\section{Montonon-Olive style electric/magnetic duality.}
\indent
The vanishing of the $\beta$ function for $N>4$ gauged supergravity  is reminiscent of the vanishing of the $\beta$ function for $N>2$ supersymmetric Yang-Mills. Indeed there is an analogous spin sum rule explanation \cite{Curtright:1981wv,Duff:1982yw}. Since the $s=1$, $s=1/2$  and $s=0$ all belong to the same adjoint representation with quadratic Casimir, $C$:
\be
-\delta^{ab}C=Tr T^aT^b
\ee
The one-loop beta is given by
\be 
\beta(total)=\Sigma_s d(s)\beta(s)
\ee
\be
\beta(s)=\frac{1}{96\pi^2}e^3(-1)^{2s}(1-12s^2)C
\ee
and hence vanishes for $N>2$ on using (\ref{sum}). This has profound implications. Let us begin with the conjecture of Montonen and Olive \cite{Montonen:1977sn,Witten:1978mh,Osborn:1979tq}, building on \cite{Goddard:1976qe}, that whereas magnetic monopoles are usually viewed as strongly coupled solitons and electric monopoles as weakly coupled elementary particles, there exists a dual description in which the magnetic monopoles are elementary particles and the electric monopoles are the solitons. This means that if we replace the coupling constant $e$ by $1/e$ and interchange the electric and magnetic charges, we obtain an equivalent theory. This in turn means that the coupling constant cannot get renormalized in perturbation theory and hence that the renormalization group $\beta$-function vanishes
\be
\beta(e)= 0 
\ee
When the $\theta$ angle is taken into account \cite{Witten:1979ey}, 
\be
S=\frac{\theta}{2\pi} +i \frac{4\pi}{e^2}
\ee
the conjecture amounts to an SL(2,Z) symmetry 
\be
S \rightarrow \frac{aS+b}{cS+d}
\ee
where $a,b,c,d $ are integers satisfying $ad-bc=1$, that acts on the electric charges $Q_e=e(m+n\theta/2\pi)
$ and magnetic charges $Q_m=n/e$.  Such a symmetry, called S-duality, would be non-perturbative since with $\theta=0$ , $a=d=0, b=-c=-1$, it reduces to $e \rightarrow 1/e$
 Strong evidence for the conjecture was provided by Sen \cite{Sen:1994fa} who pointed out that given
a purely electrically charged state $(m = 1, n = 0)$, $SL(2, Z)$ implies the existence of a state
$(p, q)$ with $p$ and $q$ relatively prime integers (i.e having no common divisor). Sen then went
on to construct explicitly a dyonic solution with charges $(1, 2)$ in complete agreement with
the conjecture. This applies to  $N = 4 $ supersymmetric Yang-Mills and also happens in certain $N = 2$
theories with vanishing beta. 
 
Several comments are now in order.

1) This S-duality conjecture was extended to string theory \cite{Font:1990gx}. Its nonperturbative nature was in contrast with the already well established T-duality arising from toroidal compactification, involving the $ radius  \rightarrow (radius)^{-1}$.  However, in six dimensions the fundamental string with coupling g admits as a soliton a dual string with coupling 1/g. On further compactification to four dimensions the non-perturbative S-duality of the fundamental string is seen to be just a perturbative T-duality of the dual string \cite{Duff:1994zt}. This D= 6 string/string duality follows in its turn from M-theory on $X^4 \times X^1$ \cite{Duff:1996rs}. The fundamental string corresponds to wrapping the M2-brane on $X^1$ to a string in D=10 then reducing to D=6 on $X^4$ whereas the dual string corresponds to wrapping the M5-brane on $X^4$ to a string in D=7 then reducing to D=6 on $X^1$. Thus in M-theory, S-duality is no longer a conjecture.  The SL(2,Z) frequently shows up as a subgroup of a larger non-compact group G known as U-duality \cite{Hull:1994ys}. Moreover, at generic points in moduli space the gauge group is abelian and the U-duality is manifestly a symmetry of the classical equations of motion. For example, ungauged N=8 supergravlty has gauge group $U(1)^{28}$ and U-duality $E_{(7,7)}$ under which the the 28 electric and 28 magnetic charges transform as a 56.

2) All this is to be contrasted with the gauged N=8  supergravity. The procees of gauging destroys the U-duality and renders the gauge group non-abelian, namely SO(8). The problem of establishing an electric/magnetic duality is thus closer to the original Montonen-Olive  set-up  where the SL(2,Z) was manifest in the mass spectrum but not in the classical equations. It seems we need someone to do for gauged supergravity what Sen did for Yang=Mills, except that not only do we lack a duality invariant mass spectrum, we  not even know what duality group we are looking for nor what spontaneous symmetry breaking of SO(8) is required, if any.

3) The Kaluza-Klein spectrum of D=11 supergravity on the round $S^7$ is purely electric and consists of a massless N=8 gauged AdS-supergravity multiplet coupled to an infinite tower of short massive AdS-supermultiplets with maximum spin 2 \cite {Duff:1983gq} . It was conjectured in 1982 \cite {Duff:1983gq} but not proved until 2012 \cite{Nicolai:2011cy} that there is a consistent truncation to the gauged  N=8 supergravity of de Wit and Nicolai \cite{deWit:1982bul}. Anti-de Sitter black hole solutions of this theory were presented in \cite{Duff:1999gh}. By focusing on the $U(1)^4$ Cartan subgroup, one finds non-extremal 1, 2, 3 and 4 charge solutions. In the extremal limit,
they may preserve up to 1/2, 1/4, 1/8 and 1/8 of the supersymmetry, respectively. In the limit of vanishing SO(8) coupling constant, the solutions reduce to the familiar black holes of the $M^4 \times T^7$  vacuum. It was  conjectured that a subset of the extreme electric black holes preserving 1/2 the supersymmetry, and which therefore form short supermultiplets, may be identified with the Kaluza-Klein states with fermion zero modes providing the spin and SO(8) quantum numbers. What about magnetic black holes? Those found in \cite{Duff:1999gh} preserved no supersymmetry but supersymmetric AdS magnetic black hole solutions maybe found elsewhere \cite{Cacciatori:2009iz,Benini:2015eyy,Ferrero:2021etw,Chamseddine:2000bk}. However they preserve fewer supersymmetries and do not seem to provide an electric/magnetic duality.

4) In related developments, it was discovered that there exists a continuous one-parameter family of inequivalent gauged SO(8) supergravities characterized by an angular parameter $\omega$ \cite{DallAgata:2012mfj,Trigiante:2016mnt}. Interestingly, the new theories allow some of the 28 gauge fields to be magnetic as well as electric. However, one can find an $\omega$-dependent transformation which renders the gauging to be purely electric in the conventional way. The theories are nevertheless still inequivalent with $\omega$-dependent potentials. 

5) We have focussed on gauged $SO(N)$ supergravity but other gaugings are possible \cite{Hull:1984qz}. In particular the authors of \cite{Guarino:2016err} look at $ISO(7)$ which follows from a consistent truncation of massive Type IIA compactified on $S^6$. The electric/magnetic deformation parameter descends directly from the Romans mass.

6) Hull \cite{Hull:2000zn} has considered strongly coupled global and local supersymmetric theories in $D=6$ with $(2,0)$ and $(4,0)$ supersymmetry, respectively, which do not admit a conventional lagrangian formulation beyond linear order \footnote{I am grateful to Leron Borsten for pointing out the zero beta connection in this context}. The first must yield $N=4$ Yang-Mills upon compactifcation on $T^2$.  What compactifications does the second allow, I wonder, and do they include AdS?

In summary, given that vanishing beta and duality go hand in hand, our conjecture is that gauged $ N>4$ supergravity also admits a Montonen-Olive style electric/magnetic duality. The electric/magnetic developments listed above are tantalising, but I have been unable to apply them to the conjecture. In particular, why only $N>4$?

The AdS/CFT version of this paper is left as an exercise for the reader.

\section{Acknowledgments}

This work is supported in part by the STFC Consolidated Grants ST/T000791/1 and ST/X000575/1
\bibliographystyle{utphys}
\bibliography{weyl.bib}

\section{Appendix:  the B coefficient }
\begin{table}[h!]
\tiny
$\begin{array}{llrrrrrrrrrrrrr}
{\cal N}=0&multiplet&f&60B&&{\cal N}=8&\\
&&&&&&&\\
\bigskip
graviton&(g_{\mu\nu})&2&-522&&1&\\
gravitino&(\psi_{\mu})&2&137&&8& \\
&&&&&&&\\
vector&(A_{\mu})&2&-12&&28&\\
&&&&&&&\\
 spinor&( \chi)&2&-3&&56\\
&&&&&&\\
scalar&(A)&1&-1&&70\\
\\\\\\
{\cal N}=1~&multiplet  &f&60B&~&  {\cal N}=8 \\
&&&&&&&\\
\bigskip
graviton&(g_{\mu\nu};\psi_{\mu})&2+2&-385&~&1&\\
gravitino&({\cal A}_{\mu}; \psi_{\mu}) &2+2&125&&~7 \\
&&&&&&&\\
vector&(\chi; A_{\mu})&2+2&-15&~&21&\\
&&&&&&&\\
 chiral &({\cal A}; \chi; A)&2+2&-5&&35&\\
&&&&&&\\
\\\\\\
{\cal N}=2~& multiplet  &f&60B&&  {\cal N}=8 &   & \\
&&&&&&&\\
\bigskip
graviton&(g_{\mu\nu}, {\cal A}_{\mu}; 2\psi_{\mu})&4+4&-260&&1
&&&&&&&\\
gravitino&({\cal A}_{\mu}; \psi_{\mu}, \chi; A_{\mu}) &4+4&110&&6& \\
&&&&&&&\\
vector&({\cal A}, 2\chi; A_{\mu}, A)&4+4&-20&&15\\
&&&&&&&\\
hyper&(2{\cal A}; 2\chi; 2A)&4+4&-10&&20&\\
\\\\\\
{\cal N}=3~  & multiplet&f&60B&&   {\cal N}=8 &
&&&&&&&\\\\
\bigskip
graviton&(g_{\mu\nu}, 2{\cal A}_{\mu};  3\psi_{\mu}, \chi;  A_{\mu})&8+8&-150&&1\\
gravitino&({\cal A}_{\mu}; \psi_{\mu}, {\cal A}: 3\chi;  2A_{\mu},A) &8+8&90&&5&\\
&&&&&&&\\
vector&(3{\cal A}; 4\chi; A_{\mu}, 3A)&8+8&-30&&10\\
\\\\\\
{\cal N}=4~  & multiplet&f&60B&&   {\cal N}=8 &   \\
&&&&&&&\\
\bigskip
graviton&(g_{\mu\nu}, 3 {\cal A}_{\mu},  {\cal A}, 4 \psi_{\mu}, 4 \chi, 3  A_{\mu}, A)&16+16&-60&&1\\
gravitino&({\cal A}_{\mu},  4{\cal A},  \psi_{\mu}, 7 \chi, 3A_{\mu},4A) &16+16&60&&4&\\
&&&&&&&\\
vector&(3{\cal A};  4 \chi; A_{\mu}, 3 A)&8+8&-30&&6&\\
\\\\\\\\
{\cal N}=5~  & multiplet&f&60B&&{\cal N}=8& \\
&&&&&&&\\
\bigskip
graviton&(g_{\mu\nu}, 4{\cal A}_{\mu}, 5{\cal A}; 5\psi_{\mu}, 11\chi: 6A_{\mu}, 5A)
&32+32&0&&1&\\
gravitino&({\cal A}_{\mu},  10{\cal A},  \psi_{\mu}, 15 \chi,  5A_{\mu}, 10A) &32+32&0&&3&\\
\\\\\\\\
{\cal N}=6~  & multiplet&f&60B&&{\cal N}=8&\\
&&&&&&&\\
\bigskip
graviton&(g_{\mu\nu}, 5{\cal A}_{\mu}, 15{\cal A}; 6\psi_{\mu}, 26\chi; 11A_{\mu},15A)
&64+64&0&&1&\\
gravitino&({\cal A}_{\mu},  9{\cal A},  \psi_{\mu}, 15 \chi,  A_{\mu\nu}, 5A_{\mu},19A) &32+32&0&&2&\\
\\\\\\\\
{\cal N}=8~  & multiplet&f&60B&&{\cal N}=8  \\
&&&&&&&&&\\
graviton&(g_{\mu\nu}, 7{\cal A}_{\mu}, 35 {\cal A}; 8\psi_{\mu}, 56\chi; 21 A_{\mu},35 A)
&128+128&0&&1&\\
\end{array}$
\caption{ The $B$ coefficient for  multiplets in an  ${\cal N}=0,1,2,3,4,5,6,8$ basis.}
\label{B}
\end{table}

 \end{document}